\documentclass[12pt,a4paper]{article} 
\usepackage[latin2] {inputenc}
\usepackage{amsmath} 
\usepackage{amsfonts} 
\usepackage{amssymb} 

\newcommand{\labelb}[1]{\label{#1}}
\newcommand{\refb}[1]{(\ref{#1})}

\newcommand{\bm}[1]{\mbox{\boldmath$#1$}}

\newcommand{\idxrangeb}[3]{\ensuremath{{#1}_{#2}\ldots{}{#1}_{#3}}}
\newcommand{\idxrange}[2]{\ensuremath{\idxrangeb{#1}{1}{#2}}}

\newcommand{\idxspc}{}

\begin{document}
\author{M. Cvitan\thanks{E-mail: mcvitan@phy.hr},
  S. Pallua\thanks{E-mail: pallua@phy.hr} and
  P. Prester\thanks{E-mail: pprester@phy.hr}\\[5mm]
  \normalsize \it Department of Theoretical Physics,\\[-1mm]
  \normalsize \it Faculty of Natural Sciences and Mathematics,\\[-1mm]
  \normalsize \it University of Zagreb,\\[-1mm]
  \normalsize \it Bijeni\v{c}ka c. 32, pp. 331, 10000 Zagreb, Croatia}

\date{\normalsize Preprint ZTF 03-04}
\title{Higher curvature Lagrangians, conformal symmetry and microscopic entropy of Killing horizons}
\maketitle

\begin{abstract}
We impose a certain class of boundary conditions on Killing horizon and show for Lagrangians with arbitrary curvature dependence that one can identify a Virasoro algebra with nontrivial central charge and calculable Hamiltonian eigenvalue. Entropy can then be calculated from Cardy formula.
\end{abstract}

\section{Introduction}
In recent years intensive research was pursued on black hole entropy and in particular on microscopic interpretation of expressions for black hole entropy obtained by methods of black hole thermodynamics. For Einstein gravity this relation is given by well known \cite{BekHaw} Bekenstein--Hawking formula.
\begin{equation}
S_{\mathrm{BH}}=\frac{A}{4\pi G}
\thickspace\textrm{,}
\end{equation} 
where $A$ denotes area of black hole horizon. Progress in understanding microscopic description of this relation was achieved using ideas of conformal symmetry near horizon. In particular Carlip \cite{Carlip99} has shown that assuming some simple boundary conditions near black hole horizon, one finds a Virasoro algebra as a subalgebra of algebra of diffeomorphisms. Treating horizon as boundary gives rise to nonvanishing central charge.  One is able to count states and to obtain entropy by the well known Cardy formula \cite{Cardy} 
\begin{equation}\labelb{eqI1}
S_{\mathrm{C}}=2\pi\sqrt{\left( \frac{c}{6} - 4\Delta_{g}\right)\left( \Delta -  \frac{c}{24} \right)  }
\thickspace\textrm{.}
\end{equation} 
Here $\Delta$ is the eigenvalue of Virasoro generator $L_0$ for the state one calculates the entropy and $\Delta_{g}$ is the smallest eigenvalue. It was shown in Einstein gravity that in this way one can reproduce Bekenstein--Hawking formula. In addition a relation between energy $\Delta$ and conformal anomaly was also obtained which reads \begin{equation}\labelb{eqc12}
\frac{c}{12}=\Delta
\thickspace\textrm{.}
\end{equation} 
This method was further developed and made more explicit and precise in \cite{ParkCarlip,Dreyer:2001py,Carlip99}. For completeness, we mention, that similar results have been obtained \cite{Solod99} by different methods, and again in the framework of conformal field theory.

Similar analysis with Carlip method was done \cite{GBII} also for Gauss--Bonnet theory given with following Lagrangian
\begin{equation}\labelb{igb}
L=
-\sum_{m=0}^{[D/2]}\lambda_m\int d^D x\sqrt{-g}\mathcal{L}_m(g)
\thickspace\textrm{,}
\end{equation}
\begin{equation}\labelb{lgbm}
\mathcal{L}_m(g)=\frac{(-1)^m}{2^{m}}
\delta_{a_{1}b_{1}\ldots{}a_{m}b_{m}}
      ^{c_{1}d_{1}\ldots{}c_{m}d_{m}}
{R^{a_{1}b_{1}}}_{c_{1}d_{1}}\cdots
{R^{a_{m}b_{m}}}_{c_{m}d_{m}}
\thickspace\textrm{,}
\end{equation}
where $\delta_{a_1\ldots{}a_k}^{b_1\ldots{}b_k}$ is totally antisymmetric product of $k$ Kronecker deltas,
normalized to take values 0 and $\pm 1$.

The result of this analysis again reproduces the classical expression \refb{eqI5} for Gauss--Bonnet entropy \cite{Visser:1993nu}.
\begin{equation}\labelb{eqI5}
S = -4\pi \sum_{m=1}^{[\frac{D}{2}]}m \lambda_m \int \hat\epsilon\; \mathcal{L}_{m-1}
\thickspace\textrm{.}
\end{equation}
It is interesting that while entropy formulas for two cases are different or in other words interaction dependent, the obtained relation \refb{eqc12} between conformal anomaly and energy is the same or interaction independent. We remind also that analysis with similar final result was done also by method of Solodukhin \cite{GBI} or with an alternative of the same method by A. Giacomini, N. Pinamonti
\cite{Giacomini:2003cg}.

It has to be stressed that previously mentioned results have been obtained using the details of Einstein or Gauss--Bonnet gravity and in particular the corresponding equations of motion. However, it is well known that classical expression for entropy is given for a much more general class of Lagrangians in fact for all diffeomorphism invariant theories in a $D$-dimensional spacetime \cite{Jacobson:vj,Iyer:1994ys}.
These Lagrangians are generally of the form
\begin{equation}\labelb{eqdefgenL}
L = L(g_{ab}, R_{abcd}, \nabla_eR_{abcd}, \dots{}, \psi, \nabla_a\psi , \dots)
\thickspace\textrm{,}
\end{equation} 
where $\psi$ denotes matter fields, and dots denote derivatives up to order $m$. 
The entropy is given with the relation 
\cite{Iyer:1994ys} 
\begin{equation}\labelb{eqS8E} 
S=-2\pi\int_{\mathcal{H}} \bm{\epsilon} E^{abcd}\eta_{ab}\eta_{cd}
\thickspace\textrm{.}
\end{equation} 
Here $\mathcal{H}$ is a cross section of the horizon, $\eta_{ab}$ denotes binormal to $\mathcal{H}$, $\hat\epsilon$ is induced volume element on $\mathcal{H}$ and
\begin{equation}\labelb{eqdefgenE}
E^{abcd} = \frac{\partial \mathcal{L}}{\partial R_{abcd}}-
\nabla_{a_1}
\frac{\partial \mathcal{L}}{\partial \nabla_{a_1} R_{abcd}}+...+(-1)^m 
\nabla_{(a_1}...\nabla_{a_m)}
\frac{\partial \mathcal{L}}{\partial \nabla_{(a_1}...\nabla_{a_m)} R_{abcd}}
\end{equation} 
The derivative in \refb{eqdefgenE} is taken with $g_{ab}$ and $\nabla_a$ fixed. It is thus tempting to investigate if microscopic interpretation of entropy using conformal field theory can be obtained for larger class of theories. 

Here we shall treat Lagrangians with arbitrary dependence on curvature
\begin{equation}\labelb{eqdefL}
L = L(g_{ab}, R_{abcd})
\thickspace\textrm{,}
\end{equation} 
where now the tensor $E^{abcd}$ takes simpler form
\begin{equation}\labelb{eqdefE}
E^{abcd}=
\frac{\partial\mathcal{L}}{\partial{R_{abcd}}}
\thickspace\textrm{.}
\end{equation} 
Following \cite{Carlip99} and assuming existence of Killing horizon with certain class of boundary conditions, we shall be able in the general case to identify near horizon a subalgebra of algebra of diffeomorphisms as a Virasoro algebra. The latter will contain central charge. In fact there will be again the same relation \refb{eqc12} between central charge $c$ and energy $\Delta$ of black hole despite the very general interaction \refb{eqdefL}. In addition, the Cardy formula will reproduce the classical result \refb{eqS8E}.
We end the introduction with the remark that just recently Frolov \cite{Frolov:2003ed}
 proposed an interesting realization of these ideas based on induced gravity.

\section{From Killing horizons to Virasoro algebra}
Due to the general character of assumed Lagrangian it is suitable to use the covariant phase approach \cite{Crnkovic,Julia:2002df,Lee:nz,Iyer:1994ys}. Here one starts from variations of Lagrangian
\begin{equation}\labelb{eqvarL}
\delta\mathbf{L}[\phi]=\mathbf{E}[\phi]\delta\phi + d\mathbf{\Theta}\left[\phi,\delta\phi\right] 
\thickspace\textrm{.}
\end{equation} 
The $(D-1)$-form $\mathbf{\Theta}$ is defined with \refb{eqvarL} and depends on $g_{ab}$, $\delta g_{ab}$ and their derivatives and is linear in $\delta g_{ab}$. It is called symplectic potential. 

It was shown \cite{Iyer:1994ys} that for Lagrangian of the type \refb{eqdefL}, that is which do not contain derivatives of Riemann tensor, the symplectic current $\mathbf{\Theta}$ is of the form.
\begin{equation}\labelb{eqII2}
\Theta_{p_{\idxspc}\idxrange{a}{n-2}}=2\epsilon_{a_{\idxspc}p_{\idxspc}\idxrange{a}{n-2}}
(E^{abcd}\nabla_d\delta g_{bc}-\nabla_dE^{abcd}\delta g_{bc})
\thickspace\textrm{,}
\end{equation}
where $E^{abcd}$ is defined in \refb{eqdefE}.

To any vector field $\xi^a$ we can associate a Noether current $(D-1)$-form
\begin{equation}
\mathbf{J}[\xi] = \mathbf{\Theta}[\phi,\mathcal{L}_\xi\phi] - \xi\cdot\mathbf{L}
\thickspace\textrm{,}
\end{equation} 
and the Noether charge $(D-2)$-form
\begin{equation}\labelb{eqJ8dQ}
\mathbf{J} = d\mathbf{Q}
\thickspace\textrm{.}
\end{equation} 
For all diffeomorphism invariant theories the Hamiltonian is a pure surface term \cite{Iyer:1994ys}
\begin{equation}\labelb{eqII7}
\delta H[\xi]=\int_{\partial C}
(\delta\mathbf{Q}[\xi]-\xi\cdot\mathbf{\Theta}[\phi,\delta\phi])
\thickspace\textrm{.}
\end{equation} 
The integrability condition requires that a $(D-1)$-form $B$ exists with the property
\begin{equation}\labelb{eqII8}
\delta\int_{\partial C}\xi\cdot\mathbf{B}=\int_{\partial C}\xi\cdot\mathbf{\Theta}
\thickspace\textrm{,}
\end{equation} 
where $C$ is a Cauchy surface.
As bulk terms of $H$ vanish, variation of $H[\xi]$ is equal to variations of boundary term $J[\xi]$. As explained in \cite{BroHen86,Carlip99} that enables one to obtain the Dirac bracket ${\left\lbrace J[\xi_1], J[\xi_2] \right\rbrace}^*$
\begin{equation}\labelb{eqII15}
{\left\lbrace J[\xi_1], J[\xi_2] \right\rbrace}^*=
\int_{\partial C} \left(
     \xi_2\cdot\mathbf{\Theta}[\phi,{\mathcal L}_{\xi_1}\phi]
   - \xi_1\cdot\mathbf{\Theta}[\phi,{\mathcal L}_{\xi_2}\phi] 
   - \xi_2\cdot(\xi_1\cdot{\bf L}) \right)
\thickspace\textrm{,}
\end{equation} 
and the algebra
\begin{equation}\labelb{eqII14}
{\left\lbrace J[\xi_1], J[\xi_2] \right\rbrace}^*
= J[\left\lbrace \xi_1, \xi_2 \right\rbrace] +
K[ \xi_1, \xi_2 ]
\thickspace\textrm{,}
\end{equation} 
with central extension $K$.

Using \refb{eqII2}, we can get a more explicit form 
\begin{eqnarray}\labelb{eqII23}
\left\lbrace J[\xi_1], J[\xi_2] \right\rbrace^*  = 
&2&\int_{\partial C}
\{
\epsilon_{a_{\idxspc}p_{\idxspc}\idxrange{a}{n-2}}
\left(
\left(
\xi_2^p E^{abcd}\nabla_d\delta_1 g_{bc}
-\xi_2^p \nabla_d E^{abcd} \delta_1 g_{bc}
\right)\right. \nonumber \\
&&
\left.
-(1\leftrightarrow{}2)
\right)
-\xi_2\cdot(\xi_1\cdot\bf{L})
\}
\thickspace\textrm{.}
\end{eqnarray} 

We shall follow Carlip and require existence of Killing horizon in a $D$-dimensional spacetime $M$ with boundary $\partial M$ such that we have a Killing vector $\chi^a$
\begin{equation}\labelb{eqIII1}
\chi^2 = g_{ab}\chi^a\chi^b = 0\thickspace\textrm{at $\partial M$}
\thickspace\textrm{.}
\end{equation} 
One defines near horizon spacelike vector $\rho_a$
\begin{equation}\labelb{eqIII2}
\nabla_a\chi^2 = -2\kappa\rho_a
\thickspace\textrm{.}
\end{equation} 
We require that variations satisfy
\begin{equation}
{\chi^a\chi^b\over\chi^2}\delta g_{ab} \rightarrow 0 , \quad
\chi^a t^b\delta g_{ab} \rightarrow 0 \qquad 
\hbox{as $\chi^2\rightarrow 0$}
\thickspace\textrm{.}
\end{equation} 
Here $\chi^a$ and $\rho_a$ are kept fixed, $t^a$ is any unit spacelike vector tangent to $\partial M$. 
One considers diffeomorphisms generated by vector fields 
\begin{equation}\labelb{eqIII8}
\xi^a=T\chi^a+R\rho^a
\thickspace\textrm{,}
\end{equation} 
Boundary conditions imply and closure of algebra requires
\begin{equation}\labelb{eqIII9}
R=\frac{1}{\kappa}\frac{\chi^2}{\rho^2}\chi^a\nabla_a T
\thickspace\textrm{,}
\qquad 
\rho^a\nabla_a T = 0
\thickspace\textrm{.}
\end{equation} 

In addition one requires for any parameter group of diffeomorphism satisfying
$\dot{T}_\alpha=\lambda_\alpha T_\alpha$ that
\begin{equation}
\int_{\partial C} \bm{\hat\epsilon}\,
   T_\alpha T_\beta \sim \delta_{\alpha+\beta}
\thickspace\textrm{.}
\end{equation} 

In order to calculate central term from \refb{eqII14} we shall use equation \refb{eqII23} where we shall integrate over $(D-2)$-dimensional surface $\mathcal{H}$ which is the intersection of Killing horizon with the Cauchy surface $C$. In addition to Killing vector $\xi^a$ we introduce other future directed null normal 
\begin{equation}
N^a = k^a -\alpha\chi^a - t^a
\thickspace\textrm{.}
\end{equation}  
where $t^a$ is tangent to $\mathcal{H}$ , and
\begin{equation}\labelb{eqII25}
k^a \equiv -\left( \chi^a - \rho^a |\chi| / \rho \right) / \chi^2
\thickspace\textrm{.}
\end{equation} 

In this way 
\begin{equation}
\epsilon_{bca_1\dots a_{n-2}} = {\hat\epsilon}_{a_1\dots a_{n-2}}
\eta_{bc} + \dots
\thickspace\textrm{,}
\end{equation} 
and
\begin{equation}\labelb{eqeta}
\eta_{ab} = 2\chi_{[b}N_{c]}=\frac{2}{|\chi{}|\rho}\rho_{[a}\chi_{b]} + t_{[a}\chi_{b]} 
\thickspace\textrm{.}
\end{equation} 
We proceed to evaluate right hand side of \refb{eqII23} by identifying the leading terms in $\chi^2$. For the first term with $E^{abcd}$ the procedure is analogous to calculation for Gauss--Bonnet \cite{GBII} special case because the details of the interaction have not been used for this term. We used only
\begin{eqnarray}
\nabla_d\delta g_{ab} 
& \equiv & \nabla_d\nabla_a\xi_b + \nabla_d\nabla_b\xi_a\nonumber\\
& = & -2 \chi_d\chi_a\chi_b      {\ddot{T} \over \chi^4} + 
       2\chi_d\chi_{(a}\rho_{b)} 
             \left(   {\dddot{T} \over {\kappa\chi^2\rho^2}} +
                {{2 \kappa \dot{T}} \over \chi^4} \right) 
\thickspace\textrm{.}
\end{eqnarray} 
We obtain the following contribution to the integrand of \refb{eqII23} coming from the first term
\begin{eqnarray}
2\xi^2_p\eta^2_{ap}E^{abcd}\nabla_d\delta g^1_{bc} &-& (1\leftrightarrow{}2)
=
\frac{1}{2}
E^{abcd}\eta_{ab}\eta_{cd} \times \nonumber \\ 
&{}&\times \left( {1 \over {\kappa}} ({T_1\dddot{T}_2}-{T_2\dddot{T}_1}) -
  {2 \kappa }(T_1 \dot{T_2} - T_2 \dot{T_1} \right)
\thickspace\textrm{.}
\end{eqnarray}
Next we have to evaluate the second term in \refb{eqII23} that is
\begin{equation}
-2\xi^2_p\nabla_dE^{abcd}\delta g^1_{bc} - (1\leftrightarrow{}2)
\thickspace\textrm{.}
\end{equation}
This term was vanishing for Gauss--Bonnet gravity due to equation of motion. In general this is not the case but for Lagrangian given with \refb{eqdefL} we want to show that near horizon it is of the order $\chi^2$. Using \refb{eqIII8} and \refb{eqeta} this term can be written
\begin{equation}
\frac{1}{\kappa \rho^2}\chi_{[a}\rho_{b]}(\alpha \chi_c + \beta \rho_c) \nabla_d E^{abcd}
\end{equation} 
with
\begin{equation}
\alpha = \frac{1}{\kappa}({\dot{T}_2\ddot{T}_1}-{\dot{T}_1\ddot{T}_2})
\thickspace\textrm{,}
\quad\textrm{}
\beta = -({T_1\ddot{T}_2}-{T_2\ddot{T}_1})
\thickspace\textrm{.}
\end{equation} 

Next step is to connect $\nabla_d E^{abcd}$ with $\nabla_\chi E^{abcd}$. We assume that ``spatial'' derivatives are $O(\chi^2)$ near horizon (Appendix A Carlip \cite{Carlip99}) thus
\begin{equation}
\nabla_d E^{abcd} = (\frac{\chi_d \nabla_\chi}{\chi^2} + \frac{\rho_d \nabla_\rho}{\rho^2})E^{abcd} + O(\chi^2)
\thickspace\textrm{.}
\end{equation}
From \refb{eqII25} we have
\begin{equation}\labelb{eqII27}
\nabla_\rho = \frac{\rho}{|\chi|} \nabla_\chi - \rho|\chi| \nabla_k
\thickspace\textrm{.}
\end{equation}
This relation together with \refb{eqIII9} requires that (see also Dreyer \cite{Dreyer:2001py})
\begin{equation}
\nabla_k T = O(\frac{1}{\chi^2})
\thickspace\textrm{.}
\end{equation}

If we apply, on the other hand, equation \refb{eqII27} to regular quantities on horizon like $R^{abcd}, E^{abcd}$ last term is $O(\chi^2)$ and
\begin{equation}
\nabla_d E^{abcd} = \frac{(\chi_d - \rho_d)}{\chi^2}\nabla_\chi E^{abcd} + O(\chi^2)
\thickspace\textrm{.}
\end{equation}
Since $\chi^a$ is Killing vector field we have ${\mathcal L}_{\chi} g_{ab} = 0$, ${\mathcal L}_{\chi} R_{abcd} = 0$, and so
\begin{equation}
{\mathcal L}_{\chi} E^{abcd} = 0
\thickspace\textrm{,}
\end{equation}
and as a consequence (using the definition of Lie derivative)
\begin{equation}
\nabla_\chi E^{abcd} = E^{fbcd}\nabla_f \chi^a + E^{afcd} \nabla_f \chi^b + E^{abfd} \nabla_f\chi^c + E^{abcf}\nabla_f \chi^d
\thickspace\textrm{.}
\end{equation}
Using
\begin{equation}
\nabla^a\chi^b = \frac{\kappa}{\chi^2}( \chi^a\rho^b-\rho^a\chi^b) + \textrm{nonleading terms}
\thickspace\textrm{.}
\end{equation}
We get
\begin{eqnarray}
\nabla_\chi E^{abcd} &=& \frac{1}{\rho^6}\chi_{[a}\rho_{b]}(\alpha \chi_c + \beta \rho_c ) (\chi_d - \rho_d )
\nonumber\\
&&(\chi_f \rho^a E^{fbcd}
-\rho_f \chi^a E^{fbcd}
-\chi_f \rho^b E^{facd}
+\rho_f \chi^b E^{facd}
\nonumber\\
&&+\chi_f \rho^c E^{fdab}
-\rho_f \chi^c E^{fdab}
+\chi_f \rho^d E^{fcba}
-\rho_f \chi^d E^{fcba}
)
\thickspace\textrm{.}
\end{eqnarray} 
When we simplify these terms we get two classes of contribution I and II. I type is e.{}g.
\begin{eqnarray}
\frac{1}{\chi^2\rho^2}\chi^e\rho^f\chi^g\rho^hE^{efgh} &=& \frac{1}{\chi^2\rho^2}\chi_{[a}\rho_{b]}\chi_{[c}\rho_{d]}E^{abcd}\nonumber\\
&=&\frac{1}{4}\eta_{ab}\eta_{cd}E^{abcd}\nonumber\\
&=&\textrm{finite}
\thickspace\textrm{.}
\end{eqnarray} 
But these terms come always in pairs of opposite signs and so they cancel. All other terms are of type II which are of the form
\begin{equation}
\frac{1}{\chi^4}\chi^a\chi^b\rho^c\rho^d E^{abcd}
\thickspace\textrm{.}
\end{equation} 
and so they vanish due to antisymmetry properties of tensor $E$.

The term $\xi_2\cdot\xi_1\cdot{\bf L}$ in \refb{eqII23} is $O(\chi^2)$ because Lagrangian is assumed to be finite on the horizon.

Finally, from \refb{eqII23}
and \refb{eqII14}
we can calculate exactly as in \cite{GBII} Noether charge
\begin{eqnarray}\labelb{eqIII41}
Q_{\idxrangeb{c}{3}{n}}
&=&
-E^{abcd}\epsilon_{a_{\idxspc}b_{\idxspc}\idxrangeb{c}{3}{n}}
\nabla_{[c}\xi_{d]}\nonumber\\
&=&
-{1 \over 2}E^{abcd}\eta_{ab}\eta_{cd}
\left( 2 \kappa T - {\ddot{T} \over {\kappa}} \right)
\hat\epsilon_{\idxrangeb{c}{3}{n}}
\thickspace\textrm{,}
\end{eqnarray}
and central charge 
\begin{eqnarray}\labelb{eqIII44}
K[\xi_1,\xi_2] &=&
-{1 \over 2}\int_{\mathcal{H}}{\hat\epsilon}_{\idxrange{a}{n-2}}
E^{abcd}\eta_{ab}\eta_{cd}
  {1 \over {\kappa}} (\dot{T}_1\ddot{T}_2-\ddot{T}_1\dot{T}_2)
\thickspace\textrm{.}
\end{eqnarray}
We stress here that one can generalise the arguments used before \cite{Carlip99,GBII} to show integrability condition
\refb{eqII8} 
and in addition that contribution $\int_{\mathcal{H}}\xi\cdot\mathbf{B}$ vanishes. This analysis again shows that boundary condition on horizon and not type of interaction are relevant for the result. Details of the argument will be published elsewhere. For integrability condition see also 
\cite{Wald:1999wa,Julia:2002df}.

From here, proceeding as in reference \cite{GBII}, we obtain for general class of Lagrangians \refb{eqdefL} Virasoro algebra with central charge 
\begin{equation}
\frac{c}{12}=\frac{\hat{A}}{8\pi}\frac{\kappa'}{\kappa}
\thickspace\textrm{,}
\end{equation}
where 
\begin{equation}\labelb{eqII39}
\hat{A} \equiv -8\pi
\int_{\mathcal{H}}{\hat\epsilon}_{\idxrange{a}{n-2}}
E^{abcd}\eta_{ab}\eta_{cd}
\thickspace\textrm{,}
\end{equation}
$\kappa'$ is period of functions which generate diffeomorphisms and $\kappa$ is surface gravity \refb{eqIII2}.

Eigenvalue of the Hamiltonian is
\begin{equation}\labelb{eqII40}
\Delta  = \frac{\kappa}{\kappa'}\hat{A}
\thickspace\textrm{,}
\end{equation}
and entropy from Cardy formula is
\begin{equation}
S = \frac{\hat{A}}{4}\sqrt{2-\left(\frac{\kappa'}{\kappa}\right)^2}
\thickspace\textrm{.}
\end{equation}

If we take for period $\kappa'$ just the period of the Euclidean black hole then in particular
\begin{equation}
S = \frac{\hat{A}}{4}
\thickspace\textrm{,}
\end{equation}
and
\begin{equation}
\frac{c}{12}=\Delta
\thickspace\textrm{.}
\end{equation}

\section{Conclusion}
We have shown in this paper that a class of boundary conditions near Killing horizon enables to identify a subalgebra of diffeomorphisms algebra as a Virasoro algebra with nontrivial central charge leading to entropy via Cardy formula. This known result in Einstein \cite{Carlip99} and Gauss--Bonnet gravity \cite{GBI,GBII} was generalized here to general class of Lagrangian with arbitrary dependence on metric and Riemann tensor. In this way one can contribute to understanding of microscopic interpretation of entropy for this Lagrangians. An interesting relation between central charge and black hole energy eigenvalue was also obtained with the same general range of validity.

We add few remarks. Approach with conformal symmetry to probe quantum mechanics seems to get contribution from various apparently independent attempts. One is the already mentioned interpretation in terms of induced gravity \cite{Frolov:2003ed}, the other is suggestion by Birmingham, Carlip, Chen \cite{Birmingham:2003wa} that analysis with quasinormal modes \cite{Hod,Dreyer} may give insight in the detailed construction of black hole states in conformal field theory. We remind that by analysis of black hole quasi-normal modes Hod obtained further arguments in support of suggestion of Bekenstein \cite{Bekenstein} that horizon area is quantized. One is tempted to guess that for generally diffeomorphism invariant Lagrangians the quantity $\hat{A}$ given with \refb{eqII39} will be quantized.

\section*{Acknowledgements}
We would like to acknowledge the financial support
under the contract No.~0119261 of Ministery of Science and Technology of Republic of
Croatia.

\end{document}